\documentclass{ws-procs975x65}

\begin{document}

\title{IN-MEDIUM HADRONIC INTERACTIONS AND THE NUCLEAR EQUATION OF STATE \\        
Talk given at the ``International Symposium on Exotic States of Nuclear Matter" \\
Catania (Italy), June 11-15, 2007
} 

\author{F. Sammarruca$^*$ }

\address{Physics Department, University of Idaho,\\
Moscow, IDAHO 83844-0903, U.S.A. \\
$^*$E-mail: fsammarr@uidaho.edu\\
}

\begin{abstract}
Microscopic studies of nuclear matter under diverse conditions of density and 
asymmetry are of great 
contemporary interest. Concerning terrestrial applications, they relate to 
future experimental facilities that will make it possible to study systems with
extreme neutron-to-proton ratio. In this talk, 
I will review recent efforts of my group aimed at exploring nuclear interactions in the medium
through the nuclear equation of state. 
The approach we take is microscopic and relativistic, with the predicted EoS properties 
derived from realistic nucleon-nucleon potentials. 
I will also discuss work in progress. Most recently, we completed a DBHF calculation of
the $\Lambda$ hyperon binding energy in nuclear matter. 
\end{abstract}


\bodymatter

\section{Introduction}\label{fs:sec1}

The properties of hadronic interactions in a dense environment 
is a problem of fundamental relevance in nuclear physics. Such properties
are typically expressed in terms of the nuclear equation of state (EoS),
a relation between energy and density (and possibly other thermodynamic
quantities) in infinite nuclear matter. 
The infinite geometry of nuclear matter eliminates surface effects and 
makes this system a clean ``laboratory" for developing and testing models of the 
nuclear force in the medium. 

The project I review in this talk is broad-scoped. We have examined several 
EoS-sensitive systems/phenomena on a consistent footing with the purpose 
of gaining a broad overview of various aspects of the EoS. We hope this will 
help identify patterns and common problems.

Our approach is microscopic, with the starting point being realistic free-space
interactions. 
In particular, we apply the Bonn B \cite{Mac89} potential                                
in the Dirac-Brueckner-Hartree-Fock
(DBHF) approach to asymmetric nuclear matter as done earlier by the Oslo group 
\cite{Oslo}. The details of the calculations have been described previously
\cite{AS03}. As it has been known for a long time, the DBHF approximation allows
a more realistic description of the saturation properties of symmetric nuclear
matter as compared with the conventional Brueckner scheme. The leading relativistic effect
characteristic of the DBHF model turns out to be a very efficient saturation 
mechanism. We recall that such correction has been shown to simulate 
a many-body effect (the ``Z-diagrams'') through mixing of positive- and 
negative-energy Dirac spinors by the scalar interaction \cite{Bro87}.               

In what follows, I will review some of our recent results and discuss 
on-going work. I stress again that these efforts belong to the 
broader context of learning more about the behavior
of the nuclear force in the medium using the EoS of infinite matter                    
(under diverse conditions of isospin and 
spin asymmetry) as an exploratory tool. 
I also emphasize the importance of fully exploiting 
empirical information, which is becoming more available   
through collisions of neutron-rich nuclei.

\section{Isospin-asymmetric nuclear matter} 
\subsection{Seeking laboratory constraints to the symmetry potential}

Reactions induced by neutron-rich nuclei can probe the isospin dependence   
of the EoS.                                                                          
Through careful analyses 
of heavy-ion collision (HIC) dynamics one can identify observables that are sensitive
to the asymmetric part of the EOS. Among those, for instance, is the neutron-to-proton ratio in 
semiperipheral collisions of asymmetric nuclei at Fermi energies \cite{Bar+05}. 

In transport models of heavy-ion collisions, particles drift in the presence 
of an average potential while undergoing two-body collisions. 
Isospin-dependent dynamics is typically included through the {\it symmetry potential} and
isospin-dependent effective cross sections.                                       
Effective cross sections (ECS) will not be discussed in this talk. However, it is worth
mentioning that they play an outstanding role for the determination of quantities
such as the nucleon mean-free path in nuclear matter, the nuclear transparency
function, and, eventually, the size of exotic nuclei. 

\begin{figure}[t]
\begin{center}
\psfig{file=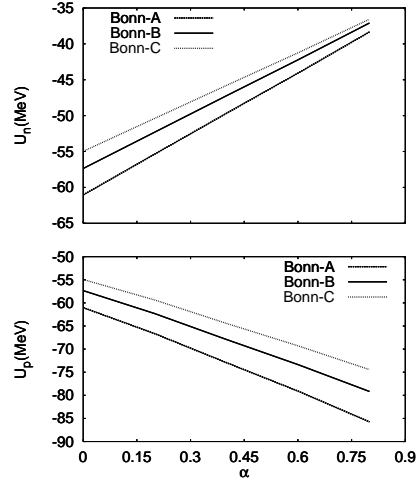,width=2in}
\end{center}
\caption{The single-neutron (upper panel) and single-proton (lower panel) potentials
as a function of the asymmetry parameter for fixed average density ($k_F=1.4
fm^{-1}$) and momentum ($k=k_F$).} 
\label{fs:fig1}
\end{figure}

\begin{figure}[t]
\begin{center}
\vspace{-2.5cm}
\psfig{file=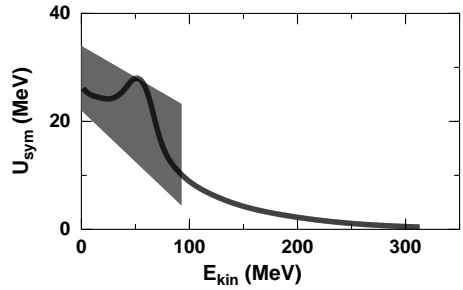,width=3.5in}
\end{center}
\vspace{-3.5cm}
\caption{The symmetry potential as a function of the nucleon kinetic energy 
at $k_F=1.4fm^{-1}$. The shaded area represents empirical information from optical
potential data.} 
\label{fs:fig2}
\end{figure}

The symmetry potential is closely related to the single-neutron/proton potentials
in asymmetric matter, which we obtain self-consistently with the effective interaction.
Those are shown in Fig.~\ref{fs:fig1} as a function of 
the asymmetry parameter $\alpha=\frac{\rho_n -\rho_p}{\rho}$. 
The approximate linear behavior, consistent with a quadratic dependence on $\alpha$
of the {\it average} potential energy, is apparent. Clearly, the isospin splitting
of the single-particle potential will be effective in separating the collision
dynamics of neutrons and protons. 
The symmetry potential is shown in Fig.~\ref{fs:fig2} where it is compared with 
empirical information on the isovector part of the optical potential (the 
early Lane potential \cite{Lane}). The decreasing strength with increasing energy
is in agreement with optical potential data. 

\subsection{Summary of EOS results and comparison with recent constraints}
\begin{figure}[t]
\begin{center}
\psfig{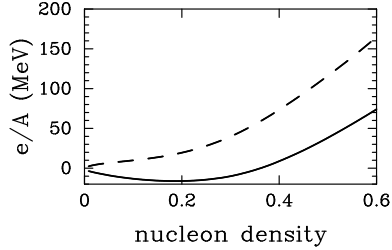}
\end{center}
\caption{Energy/particle in symmetric (solid) and neutron (dashed) matter, as a 
function of density (in fm$^{-3}$).                                                                 
} 
\label{fs:figeos}
\end{figure}

\begin{figure}[t]
\begin{center}
\psfig{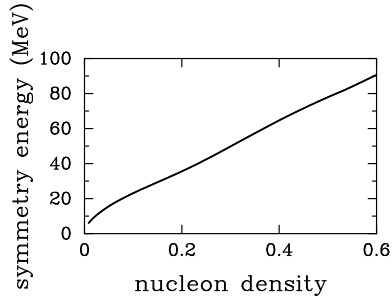}
\end{center}
\caption{Symmetry energy as a function of density                                
(in fm$^{-3}$).                                                                 
} 
\label{fs:figesym}
\end{figure}

In Table~\ref{fs:tbl1} we summarize the main properties of the symmetric and
asymmetric parts of our EoS. Those include saturation energy and density, 
incompressibility $K$, the skewness parameter $K'$, the symmetry energy, and
the symmetry pressure, $L$. The EoS for symmetric and neutron matter and the symmetry energy      
are displayed in 
Figs.~\ref{fs:figeos}-\ref{fs:figesym}.                                             

A recent analysis to constrain the EoS using compact star phenomenology and 
HIC data 
can be found in Ref.~\cite{Kl+06}.                                            
While the saturation energy is not dramatically different between models, the 
incompressibility values spread over a wider range. 
Major model dependence is found for the $K'$ parameter, where a negative value indicates
a very stiff EoS at high density. That is the case for models with parameters 
fitted to the properties of finite nuclei, whereas flow data require a soft
EoS at the higher densities and thus a larger $K'$.
 The $L$ parameter also spreads considerably, unlike the symmetry
energy which tends to be similar in most models. For $L$, a combination 
of experimental information on neutron skin thickness in nuclei and isospin 
diffusion data sets the constraint 62 MeV $< L <$ 107 MeV. 

Overall, our EoS parameters compare reasonably well with most of those constraints. They also
compare well with those from other DBHF calculations reported in Ref.~\cite{Kl+06}.

\begin{table}
\tbl{An overview of our predicted properties for the EOS of symmetric matter
and neutron matter.}
{\begin{tabular}{@{}cc@{}}
\toprule
Parameter &  Predicted Value
\\\colrule
$e_s$&        -16.14 MeV \\                                                  
$\rho_s$&     0.185 fm$^{-3}$ \\                                                
$K$ &      259 MeV        \\                                                
$K'$&      506 MeV        \\                                                
$e_{sym}(\rho_s)$ &  33.7 MeV  \\                                         
$L(\rho_s)$ &  70.1 MeV \\\botrule                                  
\end{tabular}}
\label{fs:tbl1}
\end{table}

The parameters in Table \ref{fs:tbl1} are defined by an expansion of the energy written as 
\begin{equation}
e=e_s + \frac{K}{18}\epsilon^2 - \frac{K'}{162}\epsilon^3+...+
\alpha^2(e_{sym} + \frac{L}{3}\epsilon+...), 
\label{fs:eq1} 
\end{equation}
with $\epsilon=(\rho -\rho _s)/\rho_s$, and $\rho_s$ is the saturation density.

\section{Spin-polarized neutron matter} 
In this section we move on to a another issue 
presently discussed in the literature with regard to exotic
states of nuclear/neutron matter, namely the aspect of spin asymmetry and (possible)
spin instabilities. 

The problem of spin-polarized neutron/nuclear matter has a long history. Extensive
work on the this topic has been done by Vida{\~n}a, Polls, Ramos, Bombaci, 
M{\"u}ther, and more (see bibliography of Ref.~\cite{SK07} for a more complete list).
The major driving force behind these efforts is the search for ferromagnetic
instabilities, namely the existence of a polarized state with lower energy than the
unpolarized, which naturally would lead to a spontaneous transition. 
Presently, conclusions differ widely concerning the existence of such transition
and its onset density. 

A coupled self-consistency scheme similar to the one described in Ref.~\cite{AS03}
was developed to 
calculate the EoS of polarized neutron matter. The details are described 
Ref.~\cite{SK07}. As done previously for the case of isospin asymmetry, the 
single-particle potential (for upward and downward polarized neutrons), is obtained
self-consistently with the effective interaction. Schematically
\begin{equation}
U_i = \int G_{ij} + \int G_{ii}                                       
\label{fs:eq2} 
\end{equation}
with $i,j$=$u,d$. 
The nearly linear dependence of the 
single-particle potentials on the                                               
spin-asymmetry parameter \cite{SK07}
\begin{equation}
U_{u/d}(\rho, \beta) \approx U_0(\rho, 0) \pm U_s(\rho) \beta,              
\label{fs:eq3} 
\end{equation}
with $\beta$ the spin-asymmetry parameter, 
is reminescent of the analogous case for isospin asymmetry and may be  
suggestive of a possible way to seek constraints on $U_s$, the ``spin Lane potential",
similarly to what was discussed above for the isovector optical potential. Namely, one can write,
for a nucleus,
\begin{equation}
U \approx U_0  +  U_{\sigma}({\bf s} \cdot {\bf \Sigma})/A , 
\label{fs:eq4} 
\end{equation}
with ${\bf s}$ and ${\bf \Sigma}$ the spins of the projectile nucleon and the target
nucleus, respectively, 
and extract an obvious relation with the previous equation. 
(In practice, the situation for an actual scattering experiment on a polarized 
nucleus would require a more complicated parametrization than the one above,  
as normally a spin-unsaturated nucleus is also isospin-unsaturated.)

\begin{figure}[t]
\begin{center}
\psfig{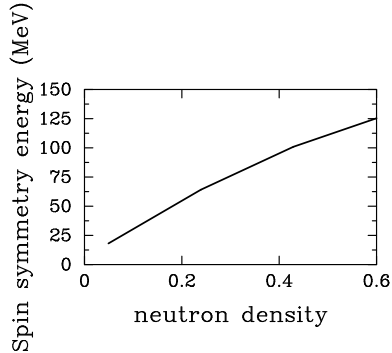}
\end{center}
\caption{Spin symmetry energy as a function of the neutron density                                
(in fm$^{-3}$).                                                                 
} 
\label{fs:figsesym}
\end{figure}

 As already implied by the linear dependence of the single-particle potential displayed
in Eq.~(\ref{fs:eq3}), 
the dependence of the average energy on the asymmetry parameter is approximately 
quadratic \cite{SK07}, 
\begin{equation}
e(\rho, \beta) \approx  e(\rho, 0) + S(\rho)\beta ^2,                       
\label{fs:eq5} 
\end{equation}
where $S(\rho)$ is the spin symmetry energy, shown in Fig.~\ref{fs:figsesym}.        
The spin symmetry energy can be related to the magnetic susceptibility
through 
\begin{equation}
\chi = \frac{\rho \mu ^2}{2 S(\rho)}.                        
\label{fs:eq6} 
\end{equation}
 The rise of $S(\rho)$ with density shows a tendency to slow down, 
a mechanism that we attribute                             
to increased repulsion in large even-singlet waves (which                
contribute only to the energy of the unpolarized state).                      
 This could be interpreted as a 
precursor of spin instability. In Table~\ref{fs:tbl2} we show predicted values of the ratio 
$\chi _F /\chi$, where $\chi _F$ is the susceptibility of a free
Fermi gas.                                                                                   

Concerning the possibility of laboratory constraints which may help shed
light on these issues, 
magnetic properties are of course closely related to the strength of the 
effective interaction in the spin-spin channel, which suggests to look into
the $G_0$ Landau parameter. With simple arguments, the latter can be related 
to the susceptibility ratio and the effective mass as 
\begin{equation}
\frac{\chi}{\chi _F} = \frac{m^*/m}{1 + G_0}                     
\label{fs:eq8} 
\end{equation}
Thus, $G_0 $$\le$ -1 signifies spin instability. (Notice the analogy with 
the formally similar relation between the incompressibility ratio $K/K_F$ and the  
parameter $F_0$, where 
$F_0 $$\le$ -1 signifies that nuclear matter is unstable.)                             
At this time, no reliable constraints on $G_0$ are available, due to the fact that
spin collective modes have not been observed with sufficient strength. 

\begin{table}
\tbl{Ratio $\chi _F /\chi$ at different densities.}                    
{\begin{tabular}{@{}cc@{}}
\toprule
$\rho /\rho _0$ & $\chi_F / \chi$ 
\\\colrule
0.5383 & 2.1455          \\                                                  
0.8886 & 2.3022          \\                                                  
1.3651 & 2.4950          \\                                                  
1.9873 & 2.6411          \\                                                  
2.7741 & 2.6824          \\                                                  
3.7457 & 2.6548          \\\botrule                                          
\end{tabular}}
\label{fs:tbl2}
\end{table}

In closing this section, we note that we found 
similar considerations concerning the trend of the magnetic susceptibility 
to hold for symmetric nuclear matter as well as for neutron matter.

\section{Work in progress: non-nucleonic degrees of freedom}
There are important motivations for considering strange baryons in nuclear matter.
The presence of hyperons in stellar matter tends to 
soften the EoS, with the consequence that the predicted             
neutron star maximum masses become considerably smaller. With recent 
constraints allowing maximum masses larger than previously accepted limits, 
accurate calculations which include strangeness become important and timely. 
On the other hand, remaining within terrestrial nuclear physics, studies 
of hyperon energies in nuclear matter naturally complements our knowledge
of finite hypernuclei.

The nucleon and the $\Lambda$ potentials in nuclear matter are the solution of a    
coupled self-consistency problem, which reads, schematically 
\begin{eqnarray}
U_N = \int_{k<k_F^N} G_{NN} + \int_{k<k_F^{\Lambda}} G_{N \Lambda}                                       
 \\ \nonumber                                              
U_{\Lambda} = \int_{k<k_F^N} G_{\Lambda N} + \int_{k<k_F^{\Lambda}} G_{\Lambda \Lambda}                             
\label{fs:eq9} 
\end{eqnarray}
To confront the simplest possible scenario, one may 
consider the case of symmetric nuclear matter at some 
Fermi momentum $k_F^N$ in the presence of a ``$\Lambda$ impurity", namely                
$k_F^{\Lambda}\approx 0$.                                                
Under these conditions, the problem stated above simplifies considerably.            
Such calculation was done in Ref.~\cite{NY94} within the Brueckner scheme. 

We have done a similar calculation but made use of              
the latest nucleon-hyperon (NY) potential of Ref.~\cite{NY05}, which was provided by 
the J{\"u}lich group.                                                                           
In a first approach, we have taken the single-nucleon potential from a separate         
calculation of symmetric matter. (Notice that the $\Lambda$ potential is
quite insensitive to the choice of $U_N$, as reported in Ref.~\cite{NY94} and as we have 
observed as well.) 
The parameters of the $\Lambda$ potential, on the other hand, are calculated self-consistently
with the $G_{N\Lambda}$ interaction, which is the solution of the Bethe-Goldstone 
equation with one-boson exchange nucleon-hyperon potentials. In the Brueckner calculation,
angle-averaged Pauli blocking and dispersive effects are included. 
Once the single-particle potential is obtained, 
 the value of $-U_{\Lambda}(p)$ at 
$p$=0 provides the $\Lambda$ binding energy in nuclear matter, $B_{\Lambda}$. 

As shown and discussed extensively in Ref.~\cite{NY05},          
there are several remarkable differences between this model and the older
$NY$ J{\"u}lich potential \cite{NY89}, and those seem to have 
a large impact on nuclear matter results. The main new feature of this model is
a microscopic model of correlated $\pi \pi$ and $K \bar{K}$ exchange to constrain both the 
$\sigma$ and $\rho$ contributions \cite{NY05}. 
With the new model, we obtain                        
considerably more attraction than Reuber {\it et al.} \cite{NY94}, about                    
49 MeV at $k_F^N$=1.35 fm$^{-1}$ for $B_{\Lambda}$.                                   
We have also incorporated the DBHF effect in this calculation (which amounts to involving the 
$\Lambda$ single-particle Dirac wave function in the self-consistent calculation through the 
effective mass) 
and find a moderate reduction of $B_{\Lambda}$ by 3-4 MeV.                              
A detailed report of this project is forthcoming. 

The natural extension of this preliminary calculation                                                 
will be a DBHF self-consistent calculation of $U_{N}$, $U_{\Lambda}$, 
and $U_{\Sigma}$ for diverse $\Lambda$ and $\Sigma$ concentrations.

\section{Summary and conclusions}
I have presented a summary of recent results from my group as well as on-going work.
Our scopes are broad and involve several aspects of nuclear matter, 
the common denominator being the behavior of the nuclear force, including
its isospin and spin dependence, in the medium. I have stressed the importance of seeking
and exploiting laboratory constraints. In the future, 
coherent effort from theory, experiment, and observations will be the key to improving 
our knowledge of nuclear matter and its exotic states.

\section*{Acknowledgments}
Support from the U.S. Department of Energy under Grant No. DE-FG02-03ER41270 is 
acknowledged. I am grateful to Johann Haidenbauer for providing the nucleon-hyperon
potential code and for useful communications.

\end{document}